\documentclass[aps,onecolumn,nobalancelastpage,nofootinbib,groupedaddress,11pt]{revtex4}

\usepackage{blindtext}
\usepackage{natbib}
\usepackage[T1]{fontenc}
\usepackage[english]{babel}
\usepackage{amsmath,amsfonts,amssymb,amsthm}
\usepackage[colorlinks=true,linkcolor=blue,urlcolor=blue,citecolor=blue]{hyperref}
\usepackage{lineno}
\usepackage{braket}
\usepackage{wrapfig}
\usepackage{todonotes}
\usepackage{graphicx,subfigure}
\usepackage{float}

\begin{document}
\title{Light propagation in inhomogeneous media, coupled quantum harmonic oscillators and phase transitions}

\author{Alejandro R. Urz\'ua}
\affiliation{Instituto Nacional de Astrof\'isica, \'Optica y Electr\'onica, Calle Luis Enrique Erro No. 1, Santa Mar\'ia Tonantzintla, Puebla, 72840, Mexico}
\email[Corresponding author: ]{arurz@inaoep.mx}
\author{Ir\'an Ramos-Prieto}
\affiliation{Instituto Nacional de Astrof\'isica, \'Optica y Electr\'onica, Calle Luis Enrique Erro No. 1, Santa Mar\'ia Tonantzintla, Puebla, 72840, Mexico}
\affiliation{Instituto de Ciencias F\'isicas, Universidad Nacional Aut\'onoma de M\'exico, Apartado Postal 48-3, 62251 Cuernavaca, Morelos, Mexico}
\author{Francisco Soto-Eguibar}
\affiliation{Instituto Nacional de Astrof\'isica, \'Optica y Electr\'onica, Calle Luis Enrique Erro No. 1, Santa Mar\'ia Tonantzintla, Puebla, 72840, Mexico}
\author{V\'ictor M. Arriz\'on}
\affiliation{Instituto Nacional de Astrof\'isica, \'Optica y Electr\'onica, Calle Luis Enrique Erro No. 1, Santa Mar\'ia Tonantzintla, Puebla, 72840, Mexico}
\author{H\'ector M. Moya-Cessa}
\affiliation{Instituto Nacional de Astrof\'isica, \'Optica y Electr\'onica, Calle Luis Enrique Erro No. 1, Santa Mar\'ia Tonantzintla, Puebla, 72840, Mexico}
\begin{abstract}
This contribution has two main purposes. First, we show using classical optics how to model two coupled quantum harmonic oscillators and two interacting quantized fields. Second, we use quantum mechanical techniques to solve, exactly, the propagation of light through a particular type of graded index medium. In passing, we show that the system presents phase transitions.
\end{abstract}

\date{\today}
\maketitle

\section{Introduction}
The existence of  analogies between quantum and classical mechanics has been applied for many years, particularly in the generation of mathematical tools to provide solutions of optical problems and vice versa \cite{nu,Nien,Anderson,Danakas,Mosh,Drago,Pare,Marte,Sch,Alonso,KVN}. The reason is that the optical paraxial wave equation is mathematically equivalent to the stationary Schr\"odinger equation, and on the other hand,  Helmholtz equation is isomorphic to the time-independent Schr\"odinger equation.\\
Classical optics systems have been used to model quantum optical phenomena \cite{Drago}. There has been also a proposal  by Man'ko {\it et al.} \cite{Mank} to realize quantum computation by using quantum like systems. Coherent random walks have been shown to occur in free propagation provided the initial  wave function is tailored properly \cite{Optica}.   The modelling of time dependent harmonic oscillators has been achieved in graded indexed media \cite{QMI} allowing to show the splitting of beams in second order solutions to the Helmholtz equation \cite{Gradient}. The propagation of classical field in a graded indexed medium with linear dependence has shown to produce and control Airy beams \cite{Airy}.\\ 
On the other hand, much attention has been recently devoted to the study of phase transitions \cite{Plenio,Jauregui} in the Rabi Model \cite{Rabi,Braak}. By means of the Holstein-Primakoff transformation \cite{Holstein}, it may be argued that two  interacting quantized fields, being, under certain approximations, a similar interaction to the Rabi interaction (approximating the spin-flip operators as creation and annihilation operators and the energy Pauli matrix as an harmonic oscillator) shows phase transitions. \\
Being aware of all of this correspondence, in this contribution we apply methods commonly used in quantum optics to solve a problem of classical optics. In particular, we show in next section that classical light propagation in a inhomogeneous medium models a coupled of quantum harmonic oscillators or the interaction of two quantized fields. Still in the same section, we show the conditions for the phase transition to occur (values of the parameters for one of the harmonic oscillators to be inverted). In section III, we show that it is possible to obtain invariant beams in the classical propagation of light in the specific  inhomogeneous medium considered. In section IV, we briefly discuss the possibility of splitting a beam by taking into account the fact that, to second order in the Helmholtz equation, a kind of quantum Kerr medium is modelled. Finally, Section V is left for conclusions.

\section{Paraxial wave equation for inhomogeneous media}
\begin{figure}
    \begin{center}
	\includegraphics[width=.55\linewidth]{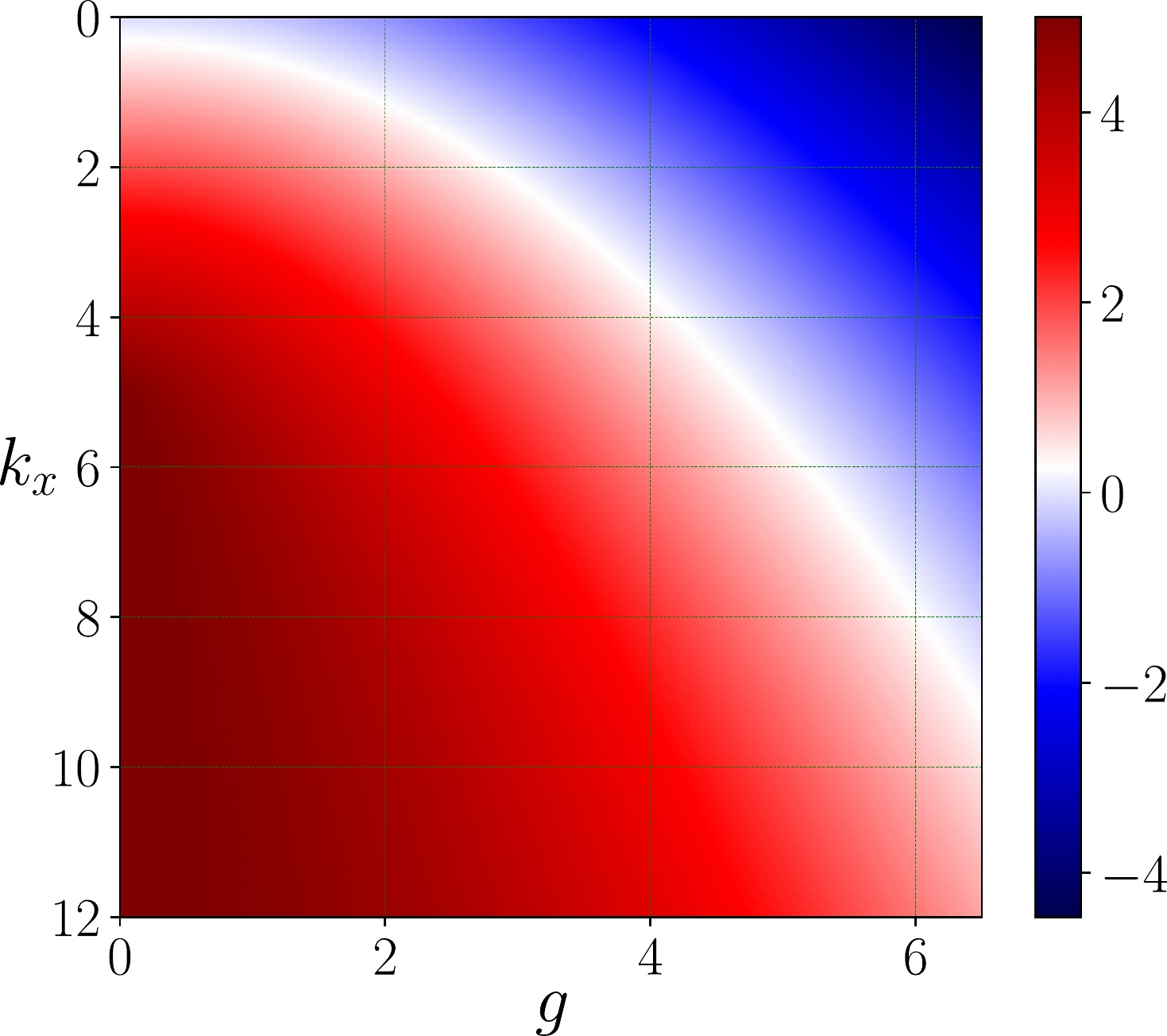}
	\caption{Plot surface of $\tilde{k}_{x}$ as a function of $k_{x}$ and $g$, maintaining $k_{y}=5$. We see that it exists a phase transition from positive to negative values of $\tilde{k}_{x}$ when $g$ goes to bigger values respect to the former.}\label{Fig_2}
	\end{center}
\end{figure}

\begin{figure}[H]
	\begin{center}
		\includegraphics[width=.55\linewidth]{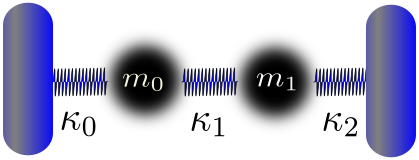}
	\caption{A lattice of two masses coupled by springs who obeys the Hamiltonian \eqref{hamiltoniano}, exhibit the same features found in the Hamiltonian \eqref{hamk} for the light propagation in an inhomogeneous medium.}\label{masas}
	\end{center}
\end{figure}

There exist classical optical systems related to waveguiding of optical beams that may be described by the paraxial wave equation
\begin{equation}\label{waveq}
2ik_{0}\frac{\partial E}{\partial z}=\nabla _{\perp }^{2}E+k^{2}(x,y)E,
\end{equation}
where $\lambda $ and $k_{0}=2\pi n_{0}/\lambda $ are the wavelength and the wavenumber of the propagating beam, respectively, and $n_{0}$ is the homogeneous refractive index. The function $k^{2}(x,y)$ describes the in-homogeneity of the medium responsible for the waveguiding of the optical field $E$. The in-homogeneity may be physical, in the sense that it is produced in the medium when fabricated; an example of such a medium is a graded index fiber. In particular, we will consider \cite{Quantum-like}
\begin{equation}\label{inhomeq}
k^{2}(x,y)=k_{0}^{2}-(k_{x}x^{2}+k_{y}y^{2})+ 2 g x y.
\end{equation}
In Reference \cite{Quantum-like}, it was proposed a way of generating the in-homogeneity function $k^2(x,y)$ by co-propagation of two beams in a Kerr nonlinear medium, that could be easier to realize experimentally. \\
Equation \eqref{waveq}, which is analogous in structure to the Schr\"odinger equation \cite{yariv}, together with the in-homogeneity \eqref{inhomeq}, has been solved for the special case of $g\ll k_x,k_y$ using the so-called rotating wave approximation \cite{Quantum-like}; here, we solve it for any set of parameters. We may rewrite Eq. \eqref{waveq} as
\begin{equation}\label{eq3}
i\frac{\partial E}{\partial z}=\hat{H} E
\end{equation}
with
\begin{equation}
\hat{H}=-\frac{1}{2 k_0}(\hat{p}_x^2+\hat{p}_y^2)+\frac{1}{k_0} \left[\frac{k_0^2}{2} -\frac{1}{2}(k_{x}x^{2}+k_{y}y^{2})+ g x y\right],
\label{hamk}
\end{equation}
where the momentum operators have the usual definition on configuration space $\hat{p}_x=-i\frac{\partial}{\partial x}$ and $\hat{p}_y=-i\frac{\partial}{\partial y}$. By considering the unitary operator
\begin{equation}\label{unitop}
\hat{R}_{\theta}=\exp[i\theta(x\hat{p}_y-y\hat{p}_x)]
\end{equation}
we make the transformation $E=\hat{R}_\theta^{\dagger}{\mathcal E}$, such that we obtain a Schr\"odinger-like equation for ${\mathcal E}$ as
\begin{equation}\label{modham}
i\frac{\partial {\mathcal E}}{\partial z}=
 \frac{1}{2k_0} \left( k_0^2  -\hat{p}_x^2 -\tilde{k}_{x}x^{2}-\hat{p}_y^2-\tilde{k}_{y}y^{2} +\tilde{g}xy\right) 
{\mathcal E},
\end{equation}
where 
\begin{subequations}\label{cf}
\begin{align}
\tilde{k}_{x}&=\frac{k_x+k_y}{2}+\frac{k_x-k_y}{2}\cos\left(2 \theta \right)-g\sin\left(  2\theta\right),\label{cfa}\\ 
\tilde{k}_{y}&=\frac{k_x+k_y}{2}-\frac{k_x-k_y}{2}\cos\left(2 \theta \right)  +g\sin\left(  2\theta\right),\label{cfb}\\ 
 \tilde{g}&=(k_x-k_y)\sin \left( 2\theta\right) +2g\cos\left(  2\theta\right),\label{cfc}
\end{align}
\end{subequations}
obtained from the set of transformations
\begin{subequations}
\begin{align}
\hat{R}_{\theta}x \hat{R}_{\theta}^{\dagger}&=x\cos\theta-y\sin\theta, \\
\hat{R}_{\theta}y \hat{R}_{\theta}^{\dagger}&=y\cos\theta+x\sin\theta, \\
\hat{R}_{\theta}\hat{p}_x \hat{R}_{\theta}^{\dagger}&=\hat{p}_x\cos\theta-\hat{p}_y\sin\theta, \\
\hat{R}_{\theta}\hat{p}_y \hat{R}_{\theta}^{\dagger}&=\hat{p}_y\cos\theta+\hat{p}_x\sin\theta.
\end{align}
\end{subequations}
From equation \eqref{cfc}, we choose $\theta=\frac{1}{2}\arctan\left(\frac{2g}{k_y-k_x}\right)$ and we have that
\begin{equation}
\cos(2\theta)=\frac{k_y-k_x}{\sqrt{(k_y-k_x)^2+4g^2}}, \qquad  \sin(2\theta)=\frac{2g}{\sqrt{(k_y-k_x)^2+4g^2}},
\end{equation}
and then equations (\ref{cf}) take the form
\begin{subequations}
\begin{align}
\tilde{k}_x&=\frac{k_x+k_y}{2} -  \frac{1}{2}\sqrt{(k_x-k_y)^2+4g^2},\label{10a}\\ 
\tilde{k}_y&=\frac{k_x+k_y}{2} +  \frac{1}{2} \sqrt{(k_x-k_y)^2+4g^2},\\
\tilde{g}&=0,
\end{align}
\end{subequations}
such that \eqref{modham} is transformed into the equation for two uncoupled quantum harmonic oscillators
\begin{equation}\label{Uncop}
i\frac{\partial {\mathcal E}}{\partial z}=
\left[  \frac{k_0}{2} - \frac{1}{2k_0} \left(\hat{p}_x^2+\tilde{k}_{x}x^{2}\right)
 - \frac{1}{2k_0} \left(\hat{p}_y^2+\tilde{k}_{y}y^{2}\right) 
\right] {\mathcal E},
\end{equation}
for which we know the formal solution.\\
We may see from equation \eqref{10a}, that for some values of the parameters $k_x, k_y, g$ of the in-homogeneity of the medium, the effective frequency of the oscillator, $\tilde{k}_x$, goes from positive to negative (see Fig.\ref{Fig_2}), therefore undergoing a phase transition \cite{Plenio,Jauregui}. It basically shows that the oscillator in the variable $x$ goes from an harmonic oscillator to an inverted harmonic oscillator, passing through a free particle \cite{Pedrosa}. In other words, we may define a critical value for $g$, 
\begin{equation}
    g_{c}=\sqrt{k_{x}k_{y}}
\end{equation}
which leads to the following three cases
\begin{equation}
\begin{array}{l}
      g<g_{c}\qquad \text{Harmonic oscillator}, \\
      g=g_{c}\qquad \text{Free particle},\\
      g>g_{c}\qquad \text{Inverted oscillator}.
\end{array}
\end{equation}
If in Equation \eqref{modham}, we write $x$ and $y$ in terms of annihilation and creation operators, we see that the so-called counter rotating terms have not been neglected (i.e., the rotating wave approximation was not performed). The presence of those terms is responsible for the phase transition in the propagation of light in the GRIN medium we are considering \cite{Quantum-like}. \\
A mechanical system equivalent to the light propagation analyzed above is given by two masses connected between them by springs and connected to a wall also by springs, as it is depicted in Fig.(\ref{masas}). The equations that rules this system is \cite{meirovitch}
\begin{subequations}\label{sis2x2}
	\begin{align}
		\kappa_1 \left(q_1-q_0\right)-\kappa_0 q_0&=m_0\ddot{q}_0,\\
		-\kappa_2q_1 -\kappa_1 \left(q_1-q_2\right)&=m_1\ddot{q}_1,
	\end{align}
\end{subequations}
where $q_0, q_1$ are the equilibrium position of mass $m_0$ and mass $m_1$, respectively, and $\kappa_0,\kappa_1,\kappa_2$ are the three spring constants. Introducing the Hamiltonian
\begin{equation}\label{hamiltoniano}
	H=\frac{p_0^2}{2 m_0}+\frac{p_1^2}{2 m_1}
	+\frac{1}{2} \kappa_0 q_0^2+\frac{1}{2} \kappa_2 q_1^2
	+\frac{1}{2} \kappa_1 \left(q_1-q_0\right)^2,
\end{equation}
in the Hamilton equations, performing some derivatives and some algebra, we get the system \eqref{sis2x2}. So, \eqref{hamiltoniano} is indeed the Hamiltonian of the system illustrated in Fig.(\ref{masas}).
We can make now the connection between our model of light propagation with this mechanical system. If in Equation \eqref{eq3} we make $z \rightarrow -t$ and
\begin{align*}
	k_0=m_0=m_1=m,\qquad
	k_x=m \frac{\kappa_0+\kappa_1}{2},\qquad
	k_y=m \frac{\kappa_1+\kappa_2}{2},\qquad
	g= m \kappa_1.
\end{align*}
the two Hamiltonians, \eqref{hamk} and \eqref{hamiltoniano}, are identified. We can go also in the other sense, from the mechanical system to the light propagation model, making $t \rightarrow -z$ and 
\begin{align*}
	m_0=m_1=m=k_0,\qquad
	\kappa_0=\frac{g-2k_x}{m},\qquad
	\kappa_1=\frac{g}{m},\qquad
	\kappa_2=\frac{g-2k_y}{m}.
\end{align*}

\section{Invariant beams}
\begin{figure}[H]
    \begin{center}
    \includegraphics[width=.55\linewidth]{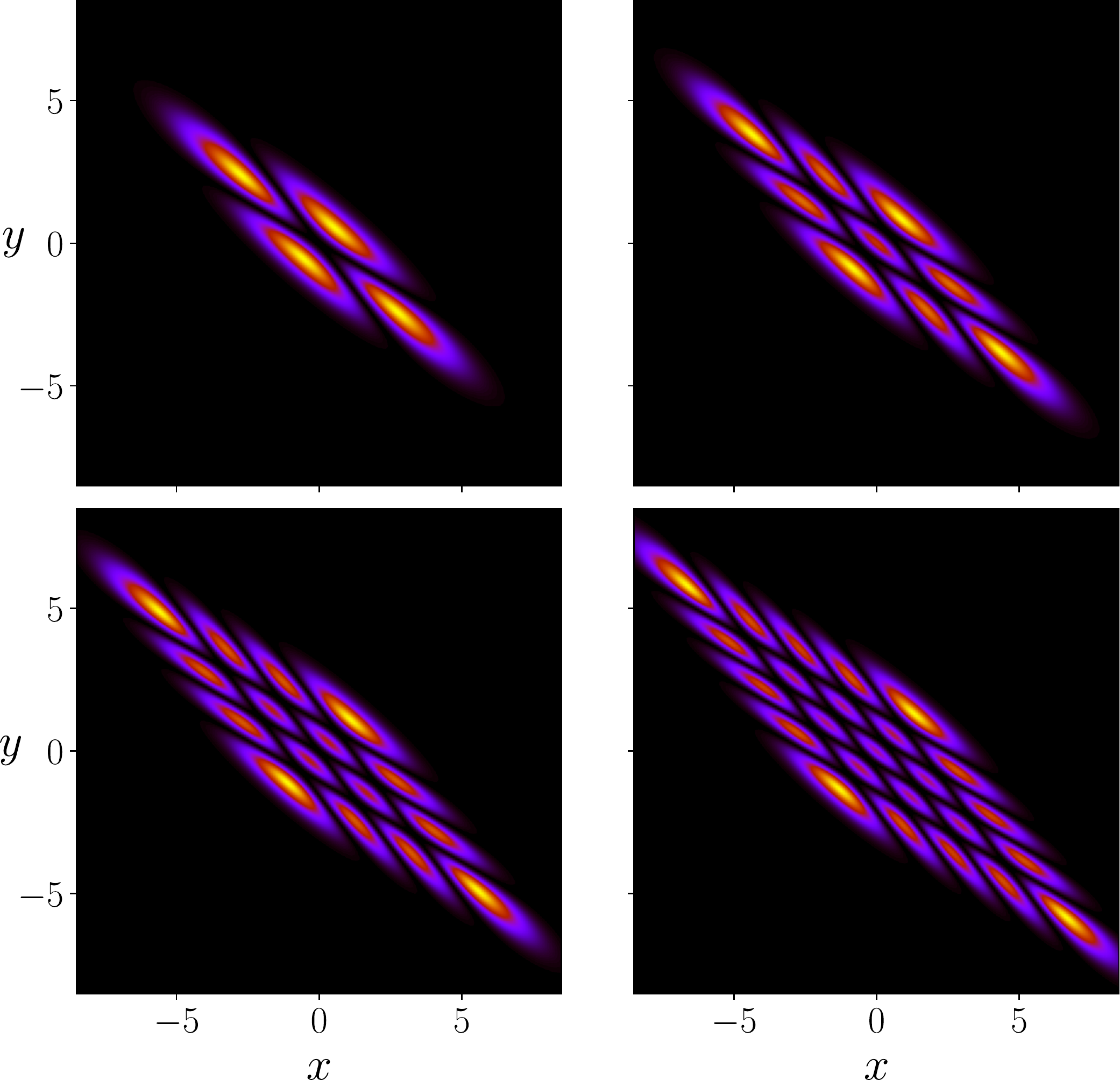}    
    \caption{Plot of the square modulus of \eqref{propa} for different values of $n_{x}$ and $n_{y}$. Upper left: $n_{x}=n_{y}=1$, upper right: $n_{x}=n_{y}=2$, lower left: $n_{x}=n_{y}=3$, lower right: $n_{x}=n_{y}=4$. For all the cases $k_{x} = 1.2$, $k_{y} = 1.5$ and $g=0.25$. We see that the propagation is invariant, i.e., it does not depend on the propagation distance $z$ despite the selection of the energy levels.}\label{Fig_3}
    \end{center}
\end{figure}

It is easy to show that there exist invariant beams for the inhomogeneity we are studying, provided that the parameter $\tilde{k}_x$ is positive such that we keep ourselves in the regime of an harmonic oscillator. Consider a field at $z=0$ given by  
\begin{equation}\label{inveq}
 E(x,y,z=0)=\hat{R}_{\theta}^{\dagger}\psi_{n_{x}}(x)\psi_{n_{y}}(y),
\end{equation} 
where $\psi_{n_{x}}(x)$ and $\psi_{n_{y}}(y)$ are Hermite-Gauss functions; i.e., eigenfunctions of the harmonic oscillators given in \eqref{Uncop}. This allows us to write the initial transformed field \eqref{inveq} as ${\mathcal E}(x,y,z=0)=\psi_{n_{x}}(x)\psi_{n_{y}}(y)$ and therefore write the solution of Equation \eqref{Uncop} as
\begin{equation}\label{UncopSol}
 {\mathcal E}(x,y,z)=
 \exp\left( i \frac{k_0 z}{2} \right) 
 \exp\left[ i \frac{z}{k_0} \left( \omega_x+\omega_y\right) \right] 
 \exp\left[ i \frac{z}{k_0} \left(n_x \omega_x+n_y \omega_y\right) \right] 
 \psi_{n_{x}}(x)\psi_{n_{y}}(y)
\end{equation}
that produces the propagated field
\begin{equation}\label{propa}
    E(x,y,z)=
    \exp\left( i \frac{k_0 z}{2} \right) 
    \exp\left[ i \frac{z}{k_0} \left( \omega_x+\omega_y\right) \right] 
    \exp\left[ i \frac{z}{k_0} \left(n_x \omega_x+n_y \omega_y\right) \right]
    \hat{R}_{\theta}^{\dagger}\psi_{n_{x}}(x)\psi_{n_{y}}(y),
\end{equation}
that is clearly an invariant beam with associated phases depending on the angular quantities $\omega_{x} = \sqrt{\tilde{k}_{x}\vphantom{\tilde{k}_{y}}}$ and $\omega_{y} = \sqrt{\tilde{k}_{y}}$, and in the independent energy levels $n_{x}$ and $n_{y}$. Some examples of the behaviour of the propagated field $E(x,y,z)$ in \eqref{propa} are shown in Fig.\ref{Fig_3}.\\
The way the operator $\hat{R}_{\theta}$ acts on the Hermite-Gauss functions on \eqref{propa} is detailed in the Appendices.  It is convenient to stress that although the propagation is invariant, there are some values of the coefficients $\{k_{x},k_{y},g\}$ for which there is a phase transition and therefore the parameter $\tilde{k}_x$ goes from positive (a usual harmonic oscillator) to negative (an inverted harmonic oscillator). In the negative region, see Fig.\ref{Fig_2}, the Hermite-Gauss functions are not anymore eigenfunctions of the inverted harmonic oscillator, therefore an invariant beam is not produced. As it was pointed out by Yuce \cite{yuce} and Mu\~noz \cite{wolf09}, the eigenfunctions of the inverted harmonic oscillator are a combination of the eigenfunctions of the free particle due to the existence of a morphism between the inverted potential and the free evolution through a proper canonical transform. Since these eigenfunctions are a combination of plane waves and waves confined in a box or propagating to a potential barrier, the non invariance argument above when $g>g_{c}$ follows.

\section{Helmholtz equation}
We consider now the complete Helmholtz equation
\begin{equation}
\frac{\partial^2E}{\partial z^2}=-\left[\frac{\partial^2}{\partial x^2}+\frac{\partial^2}{\partial y^2}+k^2(x,y)\right]E.
\end{equation}
For the case we considered in Section II, and doing the same transformation \eqref{unitop}, we arrive to
\begin{equation}
E(x,y,z)=\exp\left[ -iz
\sqrt{\frac{k_0}{2}-\frac{\omega_x}{k_0}\left( n_x+1/2\right)
-\frac{\omega_y}{k_0}\left( n_y+1/2\right) }\right] E(x,y,0),
\end{equation}
that may be developed to second order in $\kappa^2=k_0^2-k_x-k_y$ to obtain the approximation
\begin{equation}
E(x,y,z)=\exp \left[-iz\frac{\kappa}{\sqrt{2k_0}}
\left(  1-\frac{\Omega}{2\kappa^2}-\frac{\Omega^2}{8\kappa^4}\right)
 \right]E(x,y,0),
\end{equation}
where
\begin{equation}
\Omega=k_x+k_y-\omega_x\left(2n_x+1 \right)-\omega_y\left(2n_y+1 \right)  .
\end{equation}
This equation allows the splitting of fields \cite{Splitting} as the square terms resemble nonlinear quantum optical interactions, namely a quantum Kerr medium \cite{Yurke}.

\section{Conclusions}
We have shown that we can model the interaction of two masses through springs, i.e., two coupled quantum harmonic oscillators and the interaction of two quantized fields in graded index media. The solutions we have given to second order approximation of the Helmholtz equation  show how a beam splitter may be achieved in such media. We have also studied that phase transitions occur in the propagation of light in this medium, by showing that one of the harmonic oscillators is inverted for certain values of the parameters involved.

\appendix
\section{Factorization of the operator $\hat{R}_\theta$}
In this appendix the unitary operator
\begin{equation}
\hat{R}_\theta=\exp\left[ i\theta\left( \hat{x}\hat{p}_y-\hat{y}\hat{p}_x\right) \right] 
\end{equation}
is factorized.\\
We define the operators
\begin{align}
	\hat{K}_+=&\hat{x} \hat{p}_y, \\
	\hat{K}_-=&\hat{y} \hat{p}_x, \\	
	\hat{K}_0=&i\frac{1}{2}  \left(\hat{x} \hat{p}_x-\hat{y} \hat{p}_y\right).
\end{align}
The exponential of the operator $\hat{K}_0$ is nothing but a product of squeeze operators \cite{Loudon,Yuen,Caves,Vidiella,Barnett}. The above operators have the  commutators
\begin{align}
[\hat{K}_+,\hat{K}_-]=&-2\hat{K}_0, \\
[\hat{K}_0,\hat{K}_+]=&\hat{K}_+, \\	
[\hat{K}_0,\hat{K}_-]=&-\hat{K}_-,
\end{align}
and thus constitute an $su(1,1)$ algebra.\\
We propose the factorization
\begin{equation}\label{facop}
\hat{R}_\theta=\exp\left[ i\theta\left( \hat{K}_+ - \hat{K}_-\right) \right] 
=\exp\left(if_1 \hat{K}_+\right) \exp\left(if_2 \hat{K}_0\right)
\exp\left(if_3 \hat{K}_-\right)
\end{equation}
where $f_1,\;f_2,\;f_3$ are given by\\
\begin{equation}\label{f1}
f_1\left(\theta \right) = \tan\theta,
\qquad
f_2\left(\theta \right) =2i \ln\left( \cos\theta\right) ,
 \qquad
f_3\left(\theta \right) = - \tan\theta.
\end{equation}
or, writing it explicitly in terms of the original operators, 
\begin{equation}\label{rtetafinal}
\hat{R}_\theta=\exp\left[ i\theta\left( \hat{x}\hat{p}_y - \hat{y}\hat{p}_x\right) \right] 
=\exp\left[ i \tan\left( \theta\right)  \hat{x}\hat{p}_y\right]  \exp\left\lbrace  -i \ln\left[\cos\left( \theta\right)\right]   \left(\hat{x} \hat{p}_x-\hat{y} \hat{p}_y\right)\right\rbrace 
\exp\left[ -i\tan\left( \theta\right)  \hat{y}\hat{p}_x\right] .
\end{equation}

\section{Action of the operator $\hat{R}_\theta$ over a function $F\left(x,y \right) $}
It is not difficult to show that
\begin{equation}
\hat{R}_\theta F\left(x,y \right) =F\left[ \frac{\cos\left( \theta\right)x-\sin\left( \theta\right) y }{\cos^2\left( \theta\right) },
\sin\left( \theta \right) x+\cos\left( \theta \right) y  \right],
\end{equation}
where $\hat{R}_\theta $ is given in \eqref{rtetafinal} and $F(x,y)$ is an arbitrary, but well behaved, function of $x$ and $y$. \\
To study the action of the $\hat{R}_\theta $ operator over an arbitrary function $F\left(x,y \right) $, we make
\begin{equation}
\hat{R}_\theta=\exp\left[ i\theta\left( \hat{x}\hat{p}_y - \hat{y}\hat{p}_x\right) \right] 
=\hat{T}_y \hat{S}_{xy} \hat{T}_x,
\end{equation}
where
\begin{align}
\hat{T}_y=&\exp\left[ i \tan\left( \theta\right)  \hat{x}\hat{p}_y\right],
\\ 
\hat{S}_{xy}=&\exp\left\lbrace  -i \ln\left[\cos\left( \theta\right)\right],   \left(\hat{x} \hat{p}_x-\hat{y} \hat{p}_y\right)\right\rbrace 
\\
\hat{T}_x=&\exp\left[ -i\tan\left( \theta\right)  \hat{y}\hat{p}_x\right] .
\end{align}\\
Note that the operator $ \hat{S}_{xy}$ is a product of squeeze operators \cite{Loudon,Yuen,Caves,Vidiella,Barnett} in $x$ and $y$. We can prove that as
\begin{equation}\label{teo1}
\hat{T}_x F\left( x \right)=F\left[x-\tan\left( \theta \right)y  \right]  ,
\qquad
\hat{T}_y G\left( y \right)=F\left[y+\tan\left( \theta \right)x  \right] , 
\end{equation}
and the action of the squeeze operators on the variables $x$ and $y$ are
\begin{equation}
\exp\left( i r  \hat{p}_x \hat{x} \right)x=\exp(2r)x, \qquad \exp\left( i r  \hat{p}_y \hat{y} \right)y=\exp(2r)y.
\end{equation}

\end{document}